\documentstyle{europhys}

\def\etal{{\hbox{{\tenit\ et al.\/}\tenrm :\ }}}

\def\And{{\rm and\ }}

\def\stars{\bigskip\centerline{***}\medskip}

\newif\ifboo \boofalse

\def\Review#1{\boofalse{\it #1},}
\def\Name#1{{\sc #1},}
\def\Vol#1{\ifboo Vol. {\bf #1}\else{\bf #1}\fi}
\def\Year#1{\ifboo #1\else(#1)\fi}
\def\Book#1{\bootrue{\it #1},}
\def\Page#1{\ifboo {\rm p. #1}\else{\rm #1}\fi}
\input{psfig.sty}

\begin{document}

\euro{}{}{}{}
\Date{}
\shorttitle{G. ELS \etal DOPANT-BOUND SPINONS IN Cu$_{1-x}$Zn$_x$GeO$_3$}

\title{Dopant-Bound Spinons in Cu$_{1-x}$Zn$_x$GeO$_3$}

\author{G. Els\inst{1}, G.S. Uhrig\inst{2}, 
P. Lemmens\inst{1}, H. Vonberg\inst{1}, 
\\P.H.M. van Loosdrecht\inst{1}, G. G\"untherodt\inst{1}, 
\\O. Fujita\inst{3}, J. Akimitsu\inst{3},
 G. Dhalenne\inst{4}, A. Revcolevschi\inst{4}}
\institute{
\inst{1}II. Physikalisches Institut, RWTH-Aachen, Templergraben 55, 
        D-52056 Aachen, Germany.
\inst{2}Institut f\"ur Theoretische Physik, Universit\"at zu K\"oln,
        Z\"ulpicherstra\ss e 77, D-50937 K\"oln, Germany.
\inst{3}Department of Physics, Aoyama Gakuin University, 
        6-16-1 Chitsedai, Setagaya-ku, Tokyo 157, Japan.
\inst{4}Laboratoire de Chimie des Solides, Universit\'{e} 
        de Paris-Sud, b\^atiment 414, F-91405 Orsay, France.}
%\date{\today}

\rec{}{}

\pacs{
\Pacs{75}{40Gb}{Dynamic properties (susceptibility, spin waves, etc.)}
\Pacs{78}{35$+$c}{Brillouin and Rayleigh scattering}
\Pacs{75}{50Ee}{Antiferromagnetics}
      }
\maketitle

\begin{flushright}
\begin{it}
Dedicated to Prof. 
J. Zittartz,\\
on the occasion of his 60th birthday
\end{it}
\end{flushright}

\begin{abstract}
Polarized inelastic light scattering experiments on
 Cu$_{1-x}$Zn$_x$GeO$_3$ ($0\le x \le 0.045$) single crystals 
show for $x\neq0$ a new distinct mode at nearly half the energy of 
the singlet response below the spin-Peierls transition. 
The temperature, magnetic field, polarization, and doping dependencies 
of this mode are similar to those of the singlet bound state. The data 
are interpreted in terms of a spinon-assisted light scattering process. 
Position and form of the peak provide strong evidence for the presence 
of dopant-bound spinons in Cu$_{1-x}$Zn$_x$GeO$_3$. 
\end{abstract}

The discovery of the spin-Peierls (SP) transition in CuGeO$_3$ 
\cite{hase93a} led to a tremendous progress in our knowledge of 
one-dimensional antiferromagnetic Heisenberg spin-1/2 systems coupled 
to lattice degrees of freedom \cite{bouch96}. The substitution of Si 
for Ge \cite{renard95} and Zn for Cu \cite{hase93b} made it possible 
to study changes of the superexchange mechanism and of the chain 
lengths, respectively, and their effects on the SP transition temperature 
$T_{\rm SP}$ and on the singlet-triplet gap $\Delta_{\rm trip}$. 
Susceptibility \cite{renard95,hase93b} and specific heat \cite{oser95} 
experiments showed a decrease of $T_{\rm SP}$ upon increasing doping.  
Above a certain doping level the SP phase is replaced by an 
antiferromagnetically ordered phase below a N\'eel temperature $T_{\rm N}$. 
For intermediate  dopings, e.g.\ 2\% Zn-doping, a coexistence of SP and 
N\'eel phase is observed. Recent investigations indicated a coexistence 
of both phases for a broad interval of dopings 
\cite{sasa96,martin97,grenier98}.

Inelastic light scattering (ILS) is a sensitive technique to study magnetic 
excitations in pure \cite{kuroe94a,loosd96b,lemme96,loa96,loosd97b,els97} and
 doped CuGeO$_3$\cite{lemme97,sekine98a,sekine98b}. It is the aim of the present
 Letter 
to show the existence of a so far unobserved type of magnetic scattering
 process 
in doped gapful low-dimensional spin systems which arises due to the presence
 of 
dopant-bound spinons (DBS). This novel scattering process was observed in 
temperature, magnetic field, and doping-dependent polarized ILS experiments 
in the SP phase of Cu$_{1-x}$Zn$_x$GeO$_3$ for $x\neq0$.

The ILS experiments were performed, using a Sandercock-type tandem Fabry-Perot 
interferometer, as described in \cite{els97}. The temperatures $T_{\rm SP}$
 and 
$T_{\rm N}$ were checked by dc susceptibility measurements using a commercial 
SQUID magnetometer.

Fig.~\ref{dopingdep} presents ILS spectra of Cu$_{1-x}$Zn$_x$GeO$_3$ measured 
at $T$=2.2~K for various Zn concentrations. For pure CuGeO$_3$ one observes 
the well known singlet bound state (SBS), with the two-magnon continuum
 starting 
on its high frequency side. For higher temperatures a thermally activated 
excitation has been observed as a shoulder at roughly 18~cm$^{-1}$ on the low 
frequency side of the SBS \cite{els97}. The substitution of Zn for Cu 
dramatically changes the observed ILS spectra. With increasing substitution
 the 
SBS shifts to lower energies (see Fig.~\ref{dopingdep}, filled circles in lower 
inset), in agreement with the decreasing $T_{\rm SP}$. Simultaneously the SBS 
rapidly looses its intensity (see Fig.~\ref{dopingdep}, filled circles in upper 
inset)\cite{lemme96}, broadens and its shape becomes more and more symmetric. 
Additionally, a new well-defined excitation can be observed for all doped
 samples 
($x\neq0$) at nearly half the frequency of the SBS \cite{loa}. Both excitations 
show the same asymmetric line shape for low doping concentrations. On increasing 
doping the integrated peak intensity rises up to $x=1\%$; then it decreases
 up to 
$x=3.3\%$ (see Fig.~\ref{dopingdep}, open squares in upper inset). For the 
$x=4.5\%$ sample no qualitative difference is observed compared to the $x=3.3\%$ 
sample. Similar to the SBS the new mode broadens, becomes more symmetric, and 
shifts to lower energies upon increasing the doping (see Fig.~\ref{dopingdep}, 
open squares in lower inset). The new excitation is also fully (ZZ) polarized
 with 
Z$||$c-axis and shows no magnetic field dependence. Anticipating our 
interpretation 
we will attribute the new mode to a dopant-bound spinon (DBS).

Fig.~\ref{tempdep} shows the temperature dependence for the $x=0.66\%$ sample.
 At 
2.2~K the SBS at 30~cm$^{-1}$ and the mode at 15~cm$^{-1}$ can be discerned.
 Both 
peaks show a clear asymmetric line shape. On increasing the temperature up to 
$T_{\rm SP}$ both peaks display a similar behaviour. First, the integrated 
intensity 
decreases starting from a maximum value at roughly 3~K (Fig.~\ref{tempdep},
 upper 
inset). This is accompanied by a broadening of the peaks and by a symmetrisation
 of 
their shapes. Furthermore, the peak positions shift to lower frequencies 
(Fig.~\ref{tempdep}, lower inset), with a similar behaviour as in the undoped
 sample 
\cite{els97} but with reduced $T_{\rm SP}$.

Before turning to the discussion of the DBS, we first discuss several other
 possible 
origins of the new mode. The temperature dependence of the intensity of the
 new mode 
(Fig.~\ref{tempdep}, upper inset) is quite different from the one we observed
 for the 
thermally activated three-magnon process in undoped CuGeO$_3$ \cite{els97}.
 Hence we 
exclude a scattering process starting from an excited state. We also exclude
 that the 
new mode is simply due to a dopant-induced phonon scattering for two reasons.
 First, 
the scattering intensity should be more or less proportional to the doping 
concentration which is obviously not the case. Second, a dopant-induced phonon 
scattering process does not conserve momentum. Hence it should yield a rather
 broad 
feature, contrary to the observed one.

The similarities between the new mode and the SBS (same peak shapes, similar $T$ 
dependence of the peak positions and intensities (Fig.~\ref{tempdep}, insets), 
qualitatively similar doping dependence of the peak positions 
(Fig.~\ref{dopingdep}, 
lower inset) suggest that the new mode is also of magnetic origin and linked to a
bound state. The energy of the new excitation almost equals the triplet
 gap energy 
$\Delta_{\rm trip}=2.1$~meV ($\hat{=}16.8$~cm$^{-1}$) \cite{nishi94}.
 Thus one might 
think of a one-magnon process. The experiments, however, show that the DBS
 mode is 
fully (ZZ) polarized and shows neither a splitting nor a shift in magnetic 
fields. 
The interpretation for the DBS peak given previously \cite{sekine98a,sekine98b}
 is {\em not} consistent with these observations.
On these grounds,
 we exclude a simple one-magnon scattering process based on spin-orbit 
coupling  \cite{fleury68}.

The discussion of the origin of the DBS will be done in the framework of a
 simple 
model presented now. Let us view the ground state of the spin chains without
 doping 
as Resonating Valence Bond (RVB) state with singlets of various lengths. The 
dimerization which is imposed on a single chain by its adjacent chains strongly 
favors nearest-neighbour singlets on the strong bonds. Locally the ground state 
is close to a product of such singlets as in the limit of strong dimerization 
\cite{uhrig96b}. The introduction of a non-magnetic dopant like Zn breaks one of 
the singlets since one singlet partner is lacking. In this way, a magnetically
 free 
$S=1/2$ spinon is introduced \cite{martins97,hassa98}.

Doping of Si does not alter the scenario decisively. Replacing Ge by Si 
affects two adjacent chains; it is believed to cut them \cite{khoms96}.
Since the dimerization pattern alternates from chain to chain 
\cite{hirot94,brade96a} each cut breaks one weak and one strong bond.
The breaking of the strong bonds introduces two $S=1/2$ spinons
because two chain ends with weak bonds at the ends are generated
like Zn-doping introduces one chain end with a weak bond. This is
in perfect agreement with the results of Sekine {\it et al}. who observe DBS 
peaks in Zn- as in Si-doped CuGeO$_3$ samples.

Focusing on low doping concentration, we treat the spinons introduced by the
 dopants 
as independent from one another. They are, however, bound to their generating 
dopants 
\cite{lauka98}. Let us assume that the singlet pairs are at sites $(2i-1,2i)$
 for 
integer $i$ and that a dopant is present at site $1$ so that the spinon is at
 site 2. 
The spinon might move without changing the number of singlets but changing their 
position. If the spinon moves to site $2j$ there are singlets at the bonds 
$(2i',2i'+1)$
for $0 < i' < j$. For a {\em single} chain this configuration can be considered 
equivalent to the one with a spinon at site 2. For the three-dimensional 
ensemble of 
elastically coupled chains, however, the motion of the spinon costs energy which 
increases with the distance $l$ between spinon and dopant. If there is a 
sufficiently 
large next-nearest neighbour coupling $J_2$, namely 
$\alpha:=J_2/J_1 > \alpha_c\approx 0.241$ \cite{julli83,okamo92}, this 
increase is
linear in $l$ \cite{affle97}. Otherwise, it is sublinear \cite{uhrig98b}.
 Since there 
is strong evidence in CuGeO$_3$ that the frustration is above $\alpha_c$ 
\cite{riera95,fabri98} we adopt the linear potential \cite{khoms96}. For 
$\alpha>\alpha_c$ the spinons also have a finite mass and their kinetic energy 
displays a quadratic minimum in $k$-space.

The triplet excitation (``magnon'') of a dimerized, sufficiently frustrated
 chain can 
be viewed as a bound pair of two spinons coupled to $S=1$ \cite{affle97}.
 In the limit 
of small dimerization, this bound state  is described by the continuum 
Schr\"odinger 
equation
\begin{equation}
\label{trip}
\Delta_{\rm trip} \Psi(x) = -\frac{1}{2\mu} \left(\frac{\partial}{\partial x}
 \right)^2 \Psi(x) + A\delta x \Psi(x),
\end{equation}
with $x\ge 0$ being the relative coordinate and $A$ some constant. If we 
 describe the 
single spinon bound to its generating dopant in the same way we write
\begin{equation}
\label{spinon}
E_i \Psi(x) = -\frac{1}{2m} \left(\frac{\partial}{\partial x}
 \right)^2 \Psi(x) + A\delta x \Psi(x), 
\end{equation}
where $x \ge 0$ is the distance to the dopant. The only difference to 
(\ref{trip}) is 
the mass factor. For the relative motion in (\ref{trip}) we have 
$\mu^{-1} = 2m^{-1}$ 
if $m$ denotes the single spinon mass. By substituting $x \to x/2^{1/3}$ in
 (\ref{spinon}) 
we obtain $E_0 = 2^{-1/3} \Delta_{\rm trip}$ for the ground state contribution
 of the 
dopant-bound spinon. All the bound states of a linear potential are proportional
 to the 
negative zeros $-a_i$ of the Airy function \cite{abram84}. So we have for the
 first
excited level $E_1 = |a_2|/|a_1| 2^{-1/3} \Delta_{\rm trip}$, see also Fig.\ 3.
 The first 
excitation energy is
$\Delta_{\rm spinon}/E_{\rm 0} = (|a_2|-|a_1|)/|a_1|$, whence
$\Delta_{\rm spinon}/\Delta_{\rm trip} =
 2^{-1/3}(|a_2|-|a_1|)/|a_1|\approx 0.6$.  
The corresponding excitation process is visualized by the solid arrow in Fig.\ 3.
Calculations of $\Delta_{\rm spinon}/\Delta_{\rm trip}$ in more elaborate models 
will be presented elsewhere \cite{uhrig98b}.

Higher levels $E_{i\ge2}$ are not stable since they can decay into the levels 
$E_0$ or $E_1$ by producing a new pair of spinons coupled to a magnon as
 described by
eq.\ (\ref{trip}). Such a magnon is not confined to a region in space since
 it does 
not disturb the dimerization pattern other than {\em locally}. Thus the
 ladder of 
dopant-bound spinons in Fig.\ 3 is discontinued after two levels by the magnon 
continuum. The dashed arrow indicates the possibility to create unconfined
 magnons. 
The corresponding continuum yields the asymmetric tail at higher energy of
 the lines 
experimentally observed.

The question is now whether a dopant-bound spinon in the ground state $E_0$
can be excited to a state $E_1$ by a photon in a light scattering experiment. 
Such a scattering can be described by the spin conserving Raman operator which 
has the following structure 
$R= \sum_{i} (1-(-1)^i\delta) {\rm\bf S}_i {\rm\bf S}_{i+1}
 +  \gamma {\rm\bf S}_i {\rm\bf S}_{i+2}$,
where the alternation $\delta$ is the same as in the Hamiltonian, whereas the 
frustration $\gamma$ generally differs from the one in the Hamiltonian 
\cite{muthu96}. The action of $R$ on the ground state is twofold. Besides the 
spinon the usual two-magnon process takes place \cite{fleury68,els97}. For the 
action of $R$ on the spinon let us assume that the spinon is at site $i$ in the 
RVB picture. Then the relevant part of $R$ (neglecting the small alternation 
$\delta$) is 
$R_{\rm l}+R_{\rm r}$ with 
$R_{\rm l} = {\bf S}_i ({\bf S}_{i-1}+\gamma {\bf S}_{i-2})$ and
$R_{\rm r} = {\bf S}_i ({\bf S}_{i+1}+\gamma {\bf S}_{i+2})$. Applying 
$R_{\rm r}$ to the spinon at $i$ and the adjacent singlet at (i+1,i+2) yields 
(see also Ref. \cite{els97})
\begin{equation}
\label{scatter} 
R_{\rm r}|\uparrow,s\rangle = (1-\gamma)\left[(1/4) |\uparrow,t_0\rangle - 
1/(2\sqrt{2})|\downarrow,t_1\rangle \right], 
\end{equation} 
where the arrows indicate the $S_z$ component of the spinon and $s$, $t_0$ or
$t_{\pm1}$ stand for a singlet, triplet with $S_z=0$ or $S_z=\pm1$ for the two 
spins at sites $i+1$ and $i+2$. Eq.\ (\ref{scatter}) suggests that the Raman 
operator applied to the spinon creates solely an additional magnon (dashed arrow 
Fig.\ 3). Yet the state on the right hand side in (\ref{scatter}) 
is not orthogonal 
to pure spinon states. It has 75\% overlap with the state $|s,\uparrow\rangle$
where the spins at sites $i$ and $i+1$ form a singlet and the spinon is at site 
$i+2$. Thus the main  effect of $R_{\rm r}$ is to shift the spinon by one 
singlet 
spin pair to the right. In analogy $R_{\rm l}$ shifts by one singlet spin 
pair to 
the left. Thus the Raman operator is indeed able to excite the
 ground state $E_0$ 
of the dopant-bound spinon corresponding to the solid arrow in Fig.\ 3. The 
remaining 25\% effect of the action of $R$, which cannot be viewed as action 
{\em within} the single spinon subspace, are attributed to  processes like 
spinon-assisted triplet creation (dashed arrow in Fig.\ 3). But note that
the {\em total} spin of spinon {\em and} triplet remains always unchanged. 
This is in agreement with the absence of a magnetic field dependence.

From the above we conclude that the relatively sharp DBS mode visible in Figs.\ 
\ref{dopingdep} and \ref{tempdep} is due to a spinon bound to the dopant. Its 
energy is below the triplet gap $\Delta_{\rm trip}$ and its Raman line shape is 
asymmetric, since there is also a one-magnon continuum on its high
 frequency side 
belonging to it. It is possible to corroborate this picture also by 
a calculation 
in the strong dimerization limit yielding analogous results \cite{uhrig98b}.

Albeit our theoretical picture explains qualitatively very well the
experimental findings it is not yet quantitative. The experimentally observed 
sharp DBS mode is not at the theoretical value of $0.6\Delta_{\rm trip}$, but 
just below $\Delta_{\rm trip}$, and the DBS resonance and the one-magnon 
continuum are not separated. The same problem appears for the ratio of the 
singlet and the triplet gap for undoped CuGeO$_3$ \cite{bouze97a} which is 
theoretically too low ($\approx 1.5$) compared to the experimental value 
$\approx 1.85$ (cf.\ the SBS at zero doping in Fig.\ \ref{dopingdep}
 to $\Delta_{\rm trip}\approx 2.1$meV \cite{bouch96}). Binding
energies depend decisively on the dimensionality. Generically
a higher dimensionality lowers the density of states (DOS) at low 
energies. Passing from 1D to 2D to 3D, the DOS at the lower band edges 
passes from an inverse square root divergence to a steplike jump to a 
square root singularity. Thus the available phase space for the bound 
state is reduced and consequently also the binding energy  which is the 
difference of the resonance peak position from the continuum onset. Hence 
it is plausible to attribute the too low experimental binding energies to 
the higher dimensional character of the spin system in CuGeO$_3$ \cite{uhrig97a}.

For higher dopings the dopant-bound spinons start to interact by exchanging 
$S=1$ magnons. We expect that the appearance of antiferromagnetic order,
while the dimerization persists (see, e.g.,\ the shoulder close to
the laser line for $x=3.3\%$ in Fig.\ \ref{dopingdep}), can be understood 
as an ordering of these $S=1/2$ states due to their interaction 
\cite{martin97,lauka98}. But this question is far from being settled.

In summary, we performed temperature, magnetic field, polarization, and 
doping dependent inelastic light scattering experiments in
Cu$_{1-x}$Zn$_x$GeO$_3$ with $0\le x\le 4.5\%$. Besides the singlet-bound state,
 we observed a novel spinon-assisted excitation for all dopings  ($x\neq0$) 
below T$_{SP}$. We interpret this new mode as evidence for dopant-bound spinons 
which we found theoretically in an effective spinon description.

%\section{ACKNOWLEDGEMENTS}
\stars
We are grateful for helpful discussions with B.~B\"uchner, P.~Fumagalli,
 M.~Laukamp 
and F.~Sch\"onfeld. One of us (GSU) acknowledges the hospitality of the 
NHMFL, Florida. 
This work was supported by the DFG through SFB 341 and by the 
BMBF Fkz.~13N6586/8. 
Laboratoire de Chimie des Solides is ``Unit\'{e} de Recherche Associ\'{e}e au
 CNRS 
n$^{\rm o}$ 446''.

\newpage

\begin{figure}
%\begin{center}\includegraphics[width=7cm]{fig1.eps}\end{center}
%\centerline{\psfig{file=fig1.eps,width=7cm}}
\caption{
\label{dopingdep}
(ZZ) polarized ILS spectra of Cu$_{1-x}$Zn$_x$GeO$_3$ at $T=2.2~K$ 
for $0\le x\le 3.3\%$ showing the singlet bound 
state (SBS) and the dopant-bound spinon (DBS). The curves
are given an offset for clarity. The $x=0\%$ spectrum is 
reduced by $1/6$. The SBS mode at $x=0.2\%$ 
is reduced by $1/5$. The inset shows the
concentration dependence
of the integrated intensity (upper inset) and of the peak 
position (lower inset) of the SBS (filled circles) 
and the DBS (open squares).}
\end{figure}

\begin{figure}
%\begin{center}\includegraphics[width=7cm]{fig2a.eps}\end{center}
%\centerline{\psfig{file=fig2.eps,width=7cm}}
\caption{
\label{tempdep}
(ZZ) polarized ILS spectra for $x=0.66\%$ showing the temperature
 dependence of the singlet bound 
state (SBS) and the dopant-bound spinon (DBS). The curves are given an offset 
for clarity. The inset shows the temperature 
dependence of the integrated intensity (upper inset) and 
of the peak position 
(lower inset) 
of the SBS (filled 
circles) and of the DBS (open squares). Solid lines
 in the
insets are guides to the eye.}
\end{figure}

\begin{figure}
\caption{
\label{theorie}
Eigenenergies of the spinon confined to the dopant by a linear
potential.  The solid
arrow corresponds to the sharp resonance at an energy
 below the triplet gap  $\Delta_{\rm trip}$; the dashed arrow 
corresponds to the high energy tail of the asymmetric 
dopant-bound spinon (DBS) peak.}
\end{figure}

\newpage

\begin{figure}
\centerline{\psfig{file=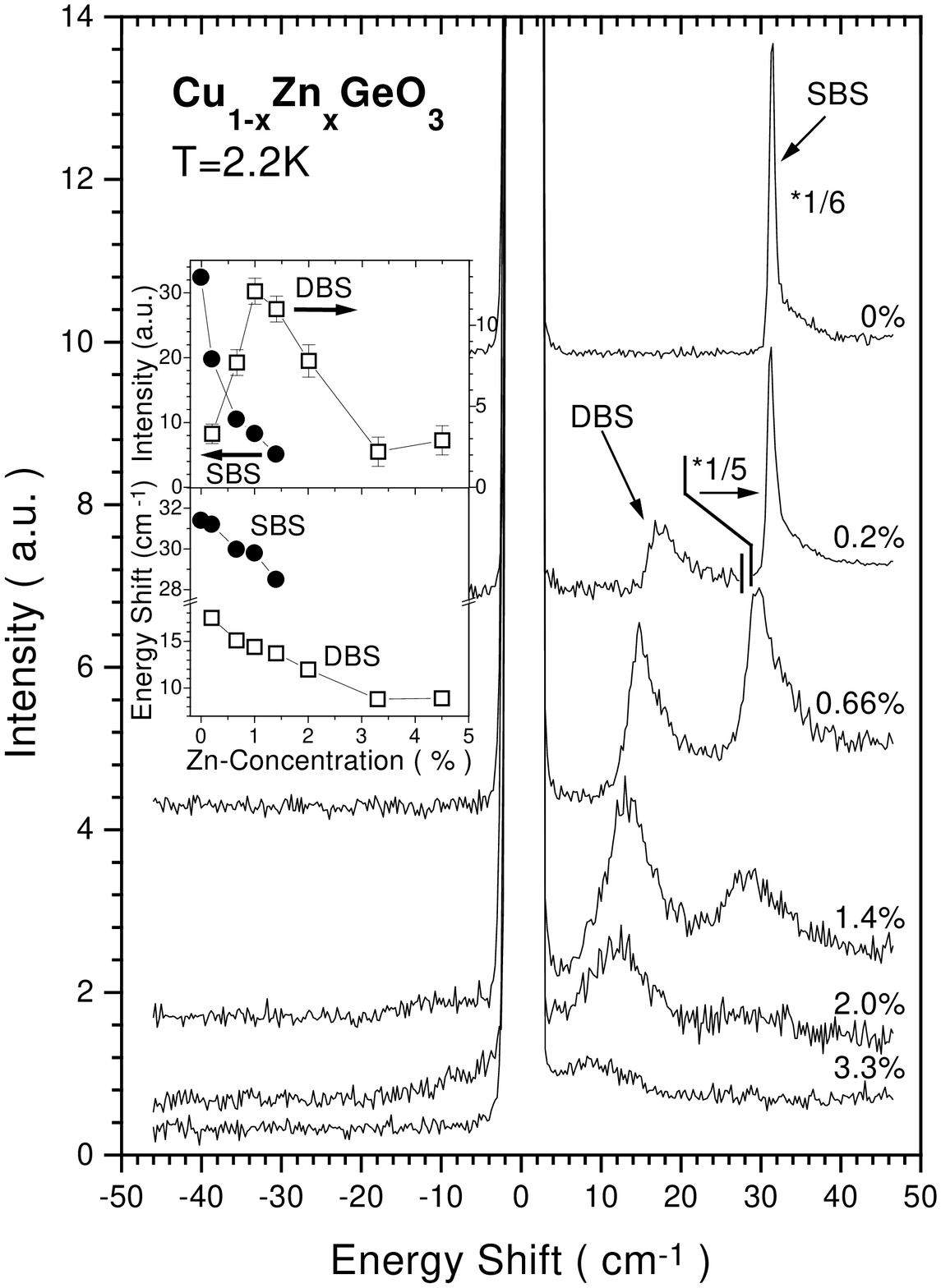,width=14cm}}
\end{figure}

\newpage

\begin{figure}
\centerline{\psfig{file=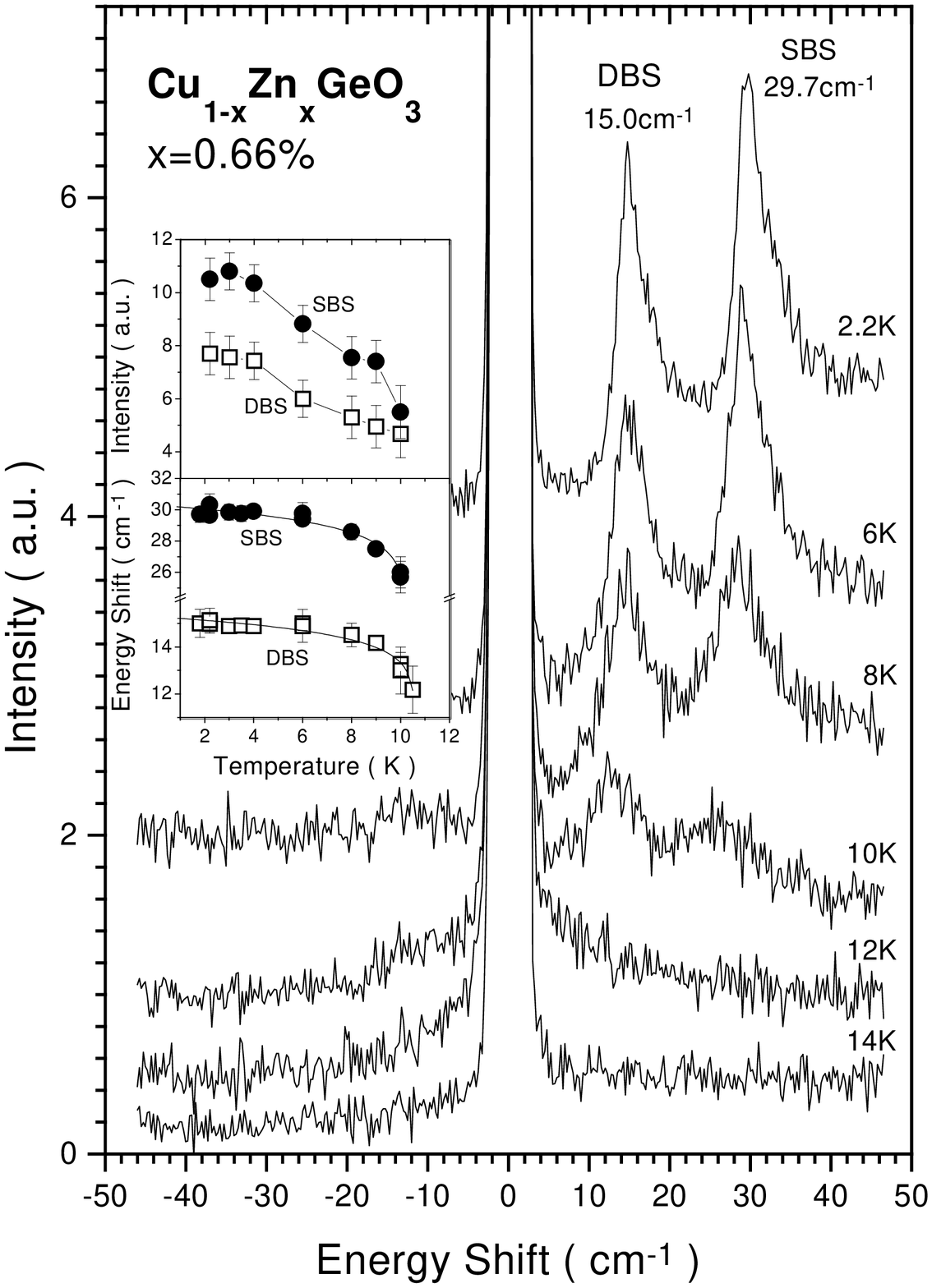,width=14cm}}
\end{figure}

\newpage

\begin{figure}
\centerline{\psfig{file=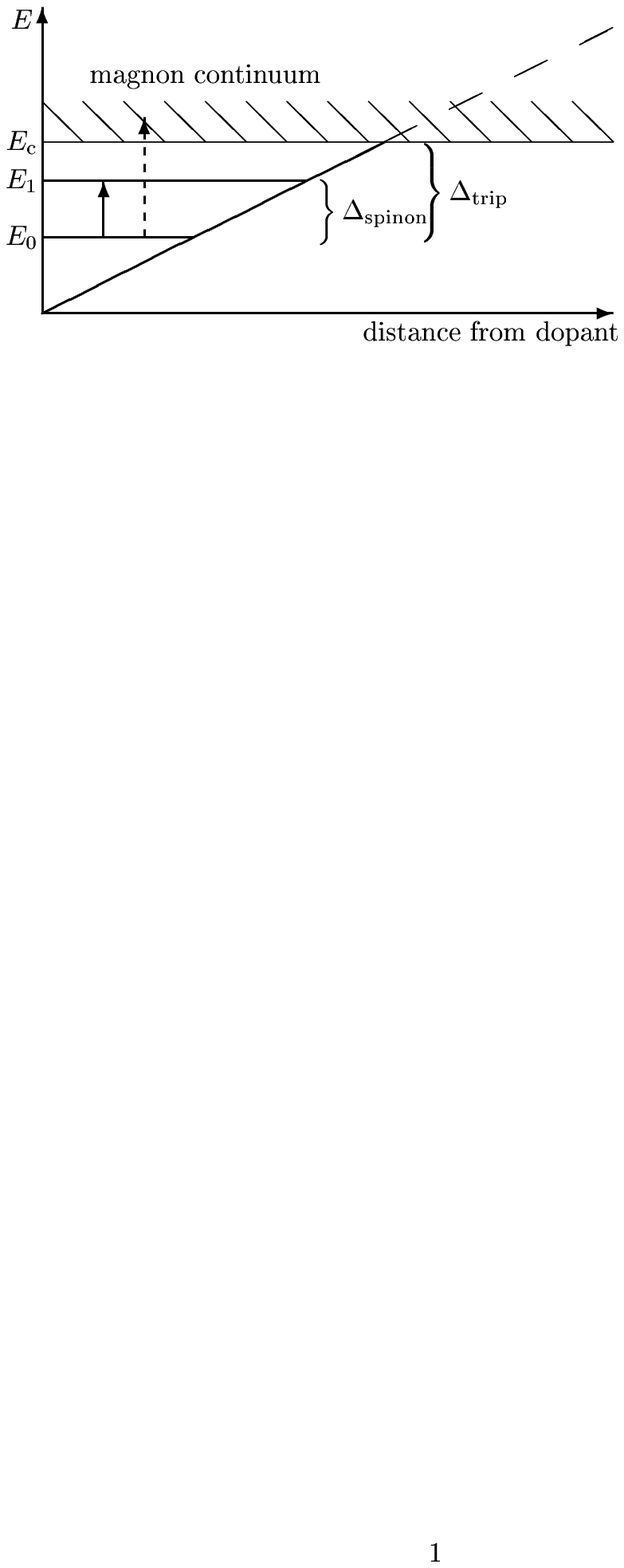,width=14cm}}
\end{figure}


\begin{thebibliography}{10}
\bibitem{loa}
The spectrum for $x=0.2\%$ (Fig.~\ref{dopingdep}) looks similar to the 
spectra 
shown in Ref. \cite{loa96} for a pure CuGeO$_3$ crystal, 
which leads to the conclusion that the 
measured samples in Ref. \cite{loa96} may have been influenced by a very 
small amount of dopants.
\end{thebibliography}
\end{document}